\documentclass[aps,pre,onecolumn,superscriptaddress]{revtex4-1}

\usepackage{amsmath}
\usepackage{times}
\usepackage{graphicx}
\usepackage{color}
\usepackage{multirow}
\usepackage{pgfplots}
\usepackage{bm}
\usepackage{ifthen}

\usepackage{bbm}
\usepackage{textcomp}
\usepackage{amssymb} 
\usepackage{amsfonts}   
\usepackage{graphicx}
\usepackage{pgfplots}
\usepackage{bm}
\usepackage{ifthen}
\newboolean{pnas}
\setboolean{pnas}{false}

\usepackage{bbm}
\usepackage{textcomp}
\usepackage{setspace}

\newcommand{\Dt}{\Delta t}

\let\a=\alpha   \let\g=\gamma  \let\d=\delta 
     \let\l=\lambda
                
 \let\t=\tau    
   
 \let\D=\Delta  \let\Th=\Theta 
         
   \let\th=\theta

\let \la = \langle \let \ra = \rangle

\begin{document}


\title{ A simple model for low variability in neural spike trains}

\author{Ulisse Ferrari}
\thanks{Correspondence should be sent to \url{ulisse.ferrari@gmail.com}.}
\affiliation{Sorbonne Universit\'e, INSERM, CNRS, Institut de la Vision, 17 rue Moreau, 75012 Paris, France.}
\author{St\'ephane Deny}
\affiliation{Neural Dynamics and Computation Lab, Stanford University, California, U.S.A}
\author{Olivier Marre}
\thanks{These authors contributed equally.}
\affiliation{Sorbonne Universit\'e, INSERM, CNRS, Institut de la Vision, 17 rue Moreau, 75012 Paris, France.}
\author{Thierry Mora}
\thanks{These authors contributed equally.}
\affiliation{Sorbonne Universit\'e, Universit\'e Paris-Diderot and \'Ecole normale sup\'erieure (PSL), 24, rue Lhomond, 75005 Paris, France}

%
\begin{abstract}
Neural noise sets a limit to information transmission in sensory systems. 
In several areas, the spiking response (to a repeated stimulus) has shown a higher degree of regularity than predicted by a Poisson process. 
However, a simple model to explain this low variability is still lacking. 
Here we introduce a new model, with a correction  to Poisson statistics, which can accurately predict the regularity of neural spike trains in response to a repeated stimulus. 
The model has only two parameters, but can reproduce the observed variability in retinal recordings in various conditions.
We show analytically why this approximation can work. 
In a model of the spike emitting process where a refractory period is assumed, we derive that our simple correction can well approximate the spike train statistics over a broad range of firing rates. 
Our model can be easily plugged to  stimulus processing models, like Linear-nonlinear model or its generalizations, to replace the Poisson spike train hypothesis that is commonly assumed. 
It estimates the amount of information transmitted much more accurately than Poisson models in retinal recordings. 
Thanks to its simplicity this model has the potential to explain low variability in other areas.
\end{abstract}

\maketitle

\bigskip

\section{Introduction}

Neural variability imposes constraints on the way neurons transmit and process information  \cite{Movshon00} and has been extensively studied in the mammalian visual system \cite{Barlow69,Heggelund78,Tolhurst81,Kara00} and beyond \cite{Baddeley97,Deweese03,Churchland10}.
The Fano Factor, i.e. the ratio of the spike count variance to the mean, is often used to characterize such variability.
Many studies have reported various values for Fano Factor depending on brain area \cite{Kara00} or experimental condition \cite{Churchland10}. 
A Fano Factor lower than one suggests that neurons spike  regularly, a condition sometimes termed as ``under-dispersion'' \cite{Gur97,Kara00,Barberini01,Deweese03,Maimon09}. 
A Fano Factor above one is termed over-dispersion \cite{Baddeley97,Churchland10}.

A common model for neural spike count is to assume Poisson statistics, where the variance in the number of spikes emitted is equal to  its mean and the Fano Factor is thus one. 
Like the classical ``LNP'' (Linear Nonlinear Poisson) \cite{Chichilnisky01}, many models describing how stimuli are processed by individual neurons rely on this assumption.
New models are therefore needed to account for the deviations from Poisson statistics observed in the data.
Several models have been proposed to account for over-dispersion in the spike count distribution \cite{Goris14,Scott12,Charles17}. 
For under-dispersed spike count distribution, a few models have been proposed, but they come with specific constraints. 
One possibility is to use a spike history filter that allows past spikes to inhibit the probability of emitting a new spike at present \cite{Pillow08}. 
However, this approach requires defining the probability of spiking over very small time bins (ex: $1ms$), and consequently needs several parameters or strong regularization to describe the spike history filter.
Other models have been proposed, but some have many parameters  (for example, $\sim 7$ for each cell in \citep{Gao15}), and may therefore require very large datasets to learn the model, or make specific assumptions that may not always be verified in the data \cite{Sellers12,Stevenson16}. 
Overall, it is unclear if a general yet simple model can account for the spike count distribution found in the data \cite{Charles17}.

Here we present a simple model that can account for under-dispersion in the spike count variability, with only two parameters. 
Our starting assumption is that the deviation from Poisson statistics comes from the refractory period, i.e. the fact that there is a minimal time interval between two spikes. 
We start from the analytic form of the spike count probability taken by a spike generation process composed of an absolute refractory period followed by a Poisson process, and show that it can be approximated by a model with only one parameter to fit to the data. 
We further simplify the model to make it amenable to log-likelihood maximization. 
We then relax our assumption by allowing for a relative refractory period \cite{Berry98,Berry98b} and we derive a more flexible model with two parameters that can accurately predict neural variability. 
The simple form of this model makes it possible to plug it on classical stimulus processing models (e.g. Linear-Nonlinear (LN) model \cite{Pillow08} or more complex cascade models \cite{Deny17}), that can be fitted with log-likelihood maximization. 
We test our model on retinal data and find that it outperforms the Poisson model, but also other statistical models proposed in the literature to account for under-dispersion. 
Either our model performs better at describing the data, or other models need more parameters for an equally accurate description. 
When combined with classical stimulus processing models, our model is able to predict the variance of the spike count over time, as well as the amount of information conveyed by single neurons, much better than classical models relying on Poisson processes.
We thus propose a simple model for neural variability, with only two parameters, that can be used in combination with any stimulus processing model to account for sub-Poisson neural variability. 


\section{Sub-Poisson behavior of retinal ganglion cells}

\begin{figure}[ht]
\centering
\includegraphics[clip=true,keepaspectratio,angle=-0,width=1.0\columnwidth]{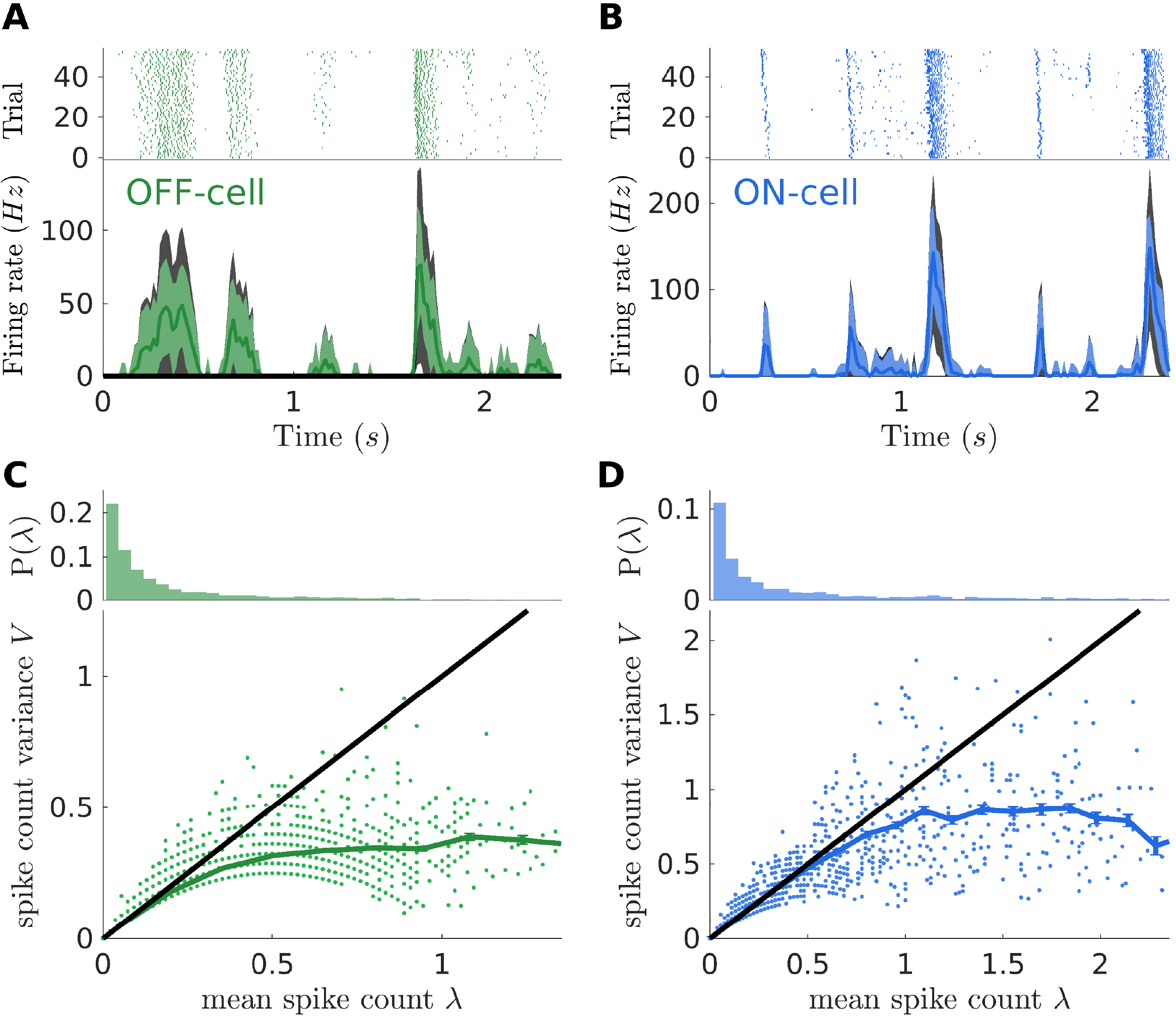}
\caption{
\textbf{Sub-Poisson behavior of RGCs.}
{\bf A}) Top:   Example raster plot for one OFF cell.
Bottom: Firing rate behavior for the same cell. Colored (resp. black) area delimit the empirical (resp. Poisson) noise, estimated as mean +/- one standard deviation.
{\bf B}) As panel A, but for an example ON cell.
{\bf C}) OFF cells show sub-Poisson behavior. Top: histogram of the observed means of the spike count across stimulus repetition (time-bin of $16.7ms$). Time-bins with zero mean have been excluded. 
Bottom: Mean of spike count across stimulus repetitions plotted against its variance. Each point correspond to one cell in one time bin. Multiple points are superimposed. Green full line: average of points with similar mean. Black line: prediction from a Poisson distributed spike count (mean equals to variance).
{\bf D}) ON cells show sub-Poisson behavior. As in C but for ON cells. Note the increase of activity with respect to OFF cells.
}
\label{fig_1}
\end{figure}
We used previously published data \citep{Deny17} where we recorded ganglion cells from the rat retina using multi-electrode arrays  \cite{Marre12,Yger16}. 
Two types of ganglion cells, ON and OFF, with receptive fields tiling the visual space, were isolated in these experiments (n$=$25 and n$=$19 cells, respectively). Cells were then stimulated with the video of two parallel horizontal bars performing a Brownian motion along the vertical direction \cite{Deny17}. Some stimuli sequences were repeated and triggered reliable responses from ganglion cells (Fig.~\ref{fig_1}A\&B top panels).

We then binned the RGC responses with $16.7ms$ time windows $\Delta T$ locked to the frame update of the video stimulation ($60Hz$).
For each cell $i$ in each time bin $t$ and for each stimulus repetition $rep$ we associated an integer spike count $n_i^{rep}(t)$ equal to the number of emitted spikes in that time-window.
In order to analyze the cell reliability as a function of the firing rate, for each cell and time bins we computed the mean ($\lambda_i(t)$) and variance ($V_i(t)$) of this spike count across repetitions of the same stimulus. 
In Fig.~\ref{fig_1}A\&B (bottom panels) we show the firing rate of two example ON and OFF cells, together with its fluctuation across the repetitions (colored areas).
For comparison we also show the fluctuations that can be expected from a Poisson process ; that is, with a variance equal to its mean.
During transients of high activity, cells are more reliable than what can be expected.

Figs.~\ref{fig_1}C\&D show that both ON and OFF cells have a sub-Poisson variability. 
Variances of spike counts are much lower than their means and their ratio (the Fano factor, equal to 1 for Poisson distributions) is not only smaller than one, but decreases with the mean.
Note that here and for the rest of the paper, we pool together all the cells of the same type together when estimating the parameters of the spike generators. 
We found that spike statistics were remarkably homogeneous across cells of the same type, and this gives us more data to fit different models. However, the same process could have been applied on a single cell, with enough repetitions of the same stimulus. 

These results are consistent with previous findings \cite{Berry98,Berry98b,Kara00}: ganglion cells emit spikes more reliably than a Poisson process. 
Consequently, a model for predicting RGC response that accounts for noise with a Poisson generator, for example the LNP model \cite{Chichilnisky01}, largely under-estimates this spiking precision. 
We then search for a simple model to account for this relation between signal and noise.


\section{Spike count statistics for a refractory neuron}
\label{refractoryModel}

Absolute or relative refractory periods are known to impose some regularity on the sequence of emitted spikes \cite{Berry98,Kara00} and thus to decrease the neural variability.
We aim at looking how refractoriness impacts the spike count distribution of the spike train. 
We consider a model of refractory neuron where the instantaneous rate is inhibited by an absolute refractory period of duration $\tau$.
Such a neuron has the following Inter-Spike interval (ISI) distribution:
\begin{equation}
\rho_\Th(t) = \Theta(t-\tau) r e^{- r ~ (t - \tau)} \label{ISI_refr}~.
\end{equation}
where $\Th(u)$ the Heaviside unit step ($\Theta(u) = 1$ for $u>0$ and zero otherwise) and $r$ is the firing rate in absence of refractoriness.
From the ISI distribution (\ref{ISI_refr}) we can estimate the probability distribution of the spike count $n$, i.e. the probability distribution of the number of spike emitted in a time bin $\Delta t$.
For the particular case of $\tau=0$, then $\lambda \equiv \la n \ra =  r \Delta t$ and $n$ follows a Poisson distribution:
\begin{equation}
P_\text{Pois}\big(n \big| \l \big) = \frac{ \l^n}{n!} e^{-\l} \label{pPois}.
\end{equation}
For $\tau > 0$, $n$ follows a sub-Poisson distribution $P_\Th(\,n\,|\,r,\Dt,\tau)$, which admits an analytic (albeit complex) expression \cite{Muller74} (see Eq.~(\ref{fullEquilRefr}) in the Methods and the derivation in the Supplementary Information).

\begin{figure}[ht]
\begin{center}
\includegraphics[clip=true,keepaspectratio,angle=-0,width=1.0\columnwidth]{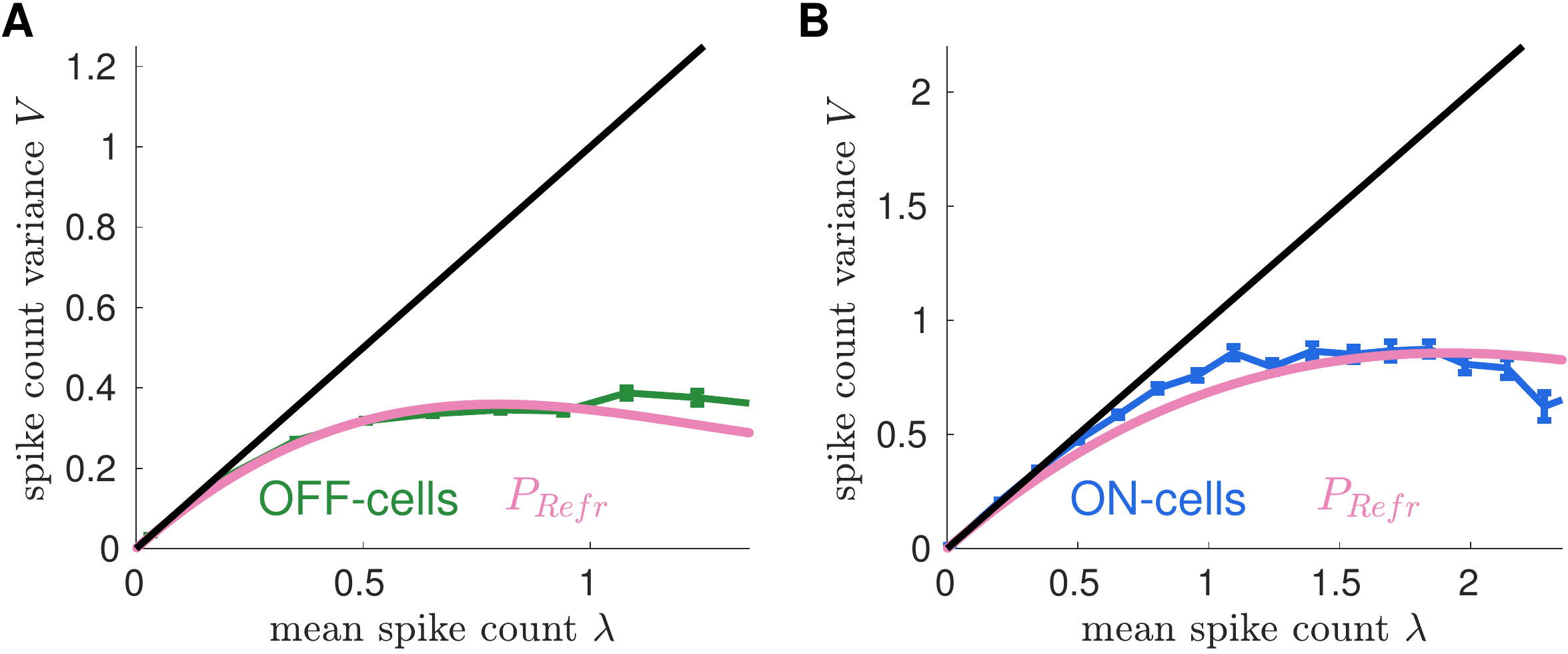}
\end{center}
\caption{\textbf{A simple neuron model with an absolute refractory period accounts for the observed spike count statistics.}
{\bf A}) Relation between mean and variance predicted by the model (pink) and measured for the OFF population (green, same data as Fig.~\ref{fig_1}).
{\bf B}) Same as A, but for the ON population.
}
\label{fig_2}
\end{figure}

To test if this expression accounts well for the data, we need to adjust the value of the refractory period $\tau$ for both cell type. 
Because the complex expression of $P_\Th(\,n\,|\,r,\Dt,\tau)$ makes it hard to apply likelihood maximization, we perform the inference of $\tau$ by minimizing the Mean-Square Error (MSE) of the mean-variance relation, summed over all  the cells and time bins of the training set (see Methods). 
We estimate $\tau$ for the two populations separately, and found $\tau = 8.8 ms$ for OFF cells and $\tau = 3.1 ms$ for ON cells.
In Fig.~\ref{fig_2} we compare the model prediction, with the mean-variance relation measured on the testing set.
The good agreement of the predictions suggests that this simple model of refractory neurons accounts well for the neural variability, even when the firing rate is large.
A simple model that takes into account the deviation from Poisson statistics due to the refractory period is thus able to predict the mean variance relation observed in the data.


\section{Simple models for the spike count statistics}
\label{modelling}

We have shown how a rather simple model of refractory neurons accurately reproduces the mean-variance relation of the recorded cells.
However, the model has a complicated analytic form, which is not easily amenable to a likelihood-maximization approach, and it cannot be considered as an alternative to models based on Poisson generators such as the LNP model.
To overcome this limitation, we propose a further simplification of the refractory neuron that is much more efficient and tractable. 

For $\tau=0$, the refractory distribution $P_\Th(\,n\,|\,r,\Dt,\tau\,)$ reduces to the Poisson distribution (Eq.~\ref{pPois}).
The small values of the refractory periods relative to the bin size suggests that an expansion at small $\tau$ could capture most of the model's behavior. 
We expand $P_\Th(\,n\,|\,r,\Dt,\tau\,)$ around the Poisson distribution to the second order in the small parameter $f=\tau/\Delta T$
(see supp. \ref{app:expansion} for details):
\begin{eqnarray}
P^\text{(2nd)}_\Th(\,n\,|\,\l,f\,) &=&  \exp\left\{ \theta_\l  n - (f-f^2) n^2 - \frac{f^2}{2} n^3 - \log n! - \log Z_\l \right\} \label{pSecond} \\
 \theta_\l &\quad& \text{such that} \quad \la n \ra_{P^\text{(2nd)}_\Th(\,n\,|\,\l,f\,)} = \lambda~.
\end{eqnarray}
The``Second-Order''  model (\ref{pSecond}) has only one free parameter, $f$ or equivalently $\tau$. $\theta_\l$ is a function of $\l$ and has to be estimated numerically in order to reproduce the mean spike count $\l$ once the refractory period $\tau$ has been fixed.  
For instance, for $\tau=0$, $\theta_\l =  \log \l$ and the Second-Order reduces to the Poisson distribution. 
Importantly, the coefficients of the $n^2$ and $n^3$ terms do not depend on the firing rate r, insofar as the refractory period does not either.
The exponential form of the model makes it easy to calculate its derivatives and to use with maximum-likelihood methods.

\begin{figure}[ht]
\begin{center}
\includegraphics[clip=true,keepaspectratio,angle=-0,width=1.0\columnwidth]{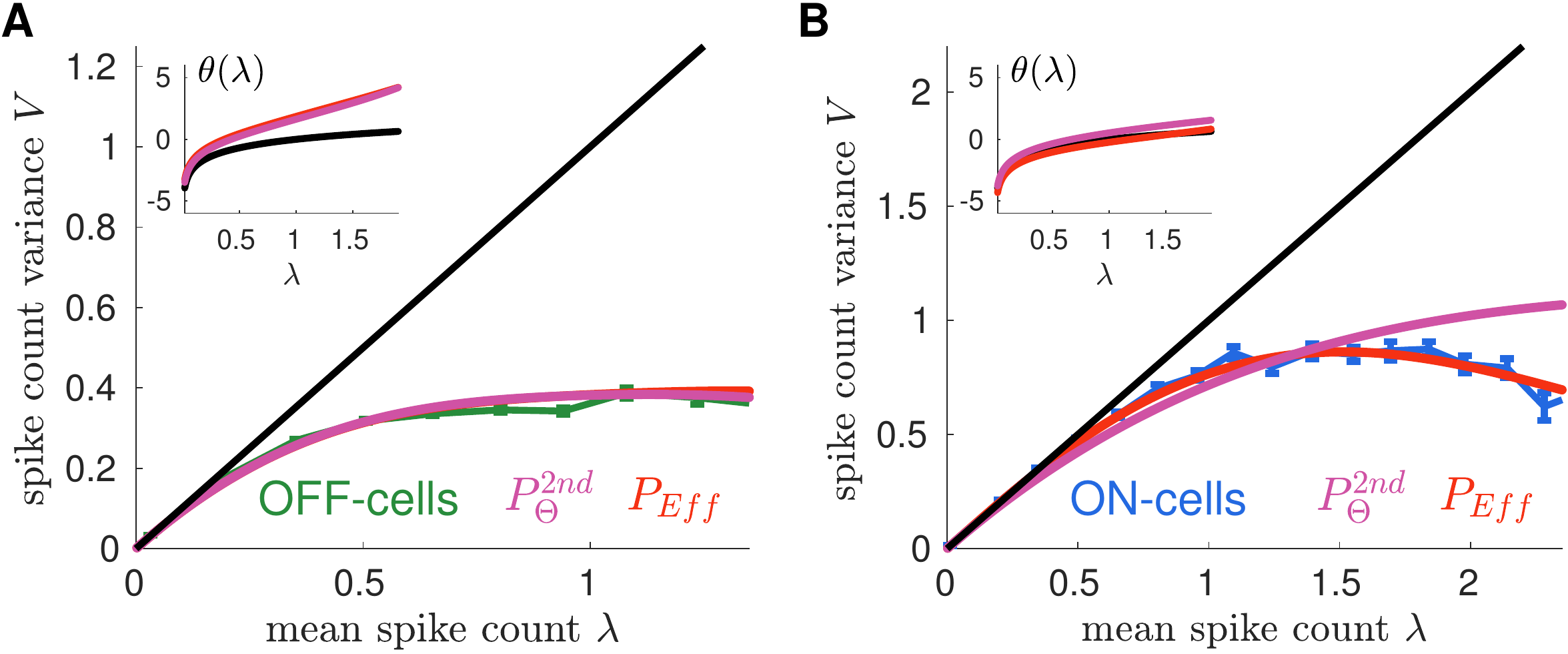}
\end{center}
\caption{\textbf{Second-Order and Effective models account well for observed spike count statistics.}
{\bf A}) The empirical behavior for the OFF population estimated from the repeated two bars stimulus (green, same data as Fig.~\ref{fig_1}) is compared with the prediction of Second-Order (purple) and Effective (red) models. Lines superimpose as the two models take a very similar form for the OFF population. For comparison, Poisson prediction is shown in black (equality line)
Inset: the function $\theta_\l$ of the two models (see text), is compared with $\theta_\l = \log \l$ for the Poisson model (black line)
{\bf B}) Same as A, but for the ON population. 
In all cases, models are learned on a separate training set.
}
\label{fig_3}
\end{figure}

We infer $\tau$ through a log-likelihood maximization (see Methods) using Eq.~\ref{pSecond}, for OFF and ON populations separately. 
From these values we can estimate the refractory period for the Second-Order model: $\tau = 10.8 ms$ for OFF cells and $\tau = 3.0 ms$ for ON cells. In Fig.~\ref{fig_3} we compare the model predictions for spike count variance $V$ with the empirical values. 
Unlike in the previous section, model parameters were not optimized to best fit this curve. 
Yet the Second-Order model shows a high performance for the OFF population, even at large firing rates. 
Despite the approximation, the model is therefore still able to describe accurately the mean variance relation, and can now be fitted using maximum likelihood approaches. 
For ON cells, the model outperforms the Poisson model but the mean variance relation is not perfectly predicted. We will then explore why this could be the case and improve our model to get a more flexible one, that can be suited for a broader range of experimental cases.

\section{A simple Effective model to describe spike trains statistics}

The previous results hold for a very specific assumption about how refractoriness constrains firing, with just a single fitting parameter -- the refractory period. We thus wondered whether more general rules of refractoriness could give rise to a broader class of sub-Poisson spiking distributions, allowing for a better agreement with the data.

We considered a general model in which the instantaneous spike rate is now inhibited by a time-dependent factor, $\alpha(r,u)r$, where $r$ is the spike rate in absence of refractoriness, and $u$ is the time following the previous spike, with $\alpha(r,0)=0$ and $\alpha(r,u\geq \tau)=1$. Calculating the distribution of number spikes under this assumption is intractable analytically, but a Second-Order expansion such as the one performed in the previous section can still be performed, and yields the following expression at Second-Order (see Supp \ref{app:expansion}):

\begin{eqnarray}
&&P^\text{(2nd)}_\a(\,n\,|\,\l,r,f,\Dt\,) =  \exp\left\{ \theta_\l  n - \g_\a n^2 - \d_\a n^3 - \log n! - \log Z_\l \right\} \label{pOfNalpha}\\
&&\quad \g_\a =   f - f^2  + (2+ r \Dt) \left(\frac{f^2}{2}- g \right) \nonumber \\
&&\quad \d_\a =   f^2 - g = \frac{f^2}{2} + \left(\frac{f^2}{2}- g \right) \nonumber \\
&&\quad \theta_\l \quad \text{such that} \quad \la n \ra_{P^\text{(2nd)}_\a(\,n\,|\,\l,r,f,\Dt\,)} = \lambda~.
\end{eqnarray}

where $f \equiv \int_0^\infty du \,(1- \alpha(r,u)) / \Dt $ and $g \equiv  \int_0^\infty du\, u\,(1-\alpha(r,u))/\Dt^2$. 
The special case of a pure refractory period, $\alpha(r,u)=0$ for $0\leq u\leq \tau$, gives back Eq.~\ref{pSecond} and the previous definition of $f$.
In this case, $f^2/2 -g$ vanishes. 
This quantity can thus be considered as an estimate of the deviation from the absolute refractoriness.
Again, $\theta_\l$ should be adjusted to match the average number spikes in each cell and time bin, $\lambda_i(t)$.

This analytic development shows that the coefficients of $n^2$ and $n^3$, $\g_\a$ and $\d_\a$, can have very different forms depending on the exact form of the refractoriness. We thus decided to relax the assumption of a strict dependence between $\g_\a$ and $\d_\a$.  
We tested if a model with $\g_\a$ and $\d_\a$ that do not depend on $r$ shows a good agreement with the data.
In the following we thus treat the coefficients of $n^2$ and $n^3$ in Eq.~(\ref{pOfNalpha} as two parameters that are independent of each other, and also independent of the firing rate.

The resulting Effective model is:
\begin{eqnarray}
P_\text{Eff}(\,n\,|\, \l,\g,\d\,) &=&  \exp\left\{ \theta_\l  n - \g n^2 - \d n^3 - \log n! - \log Z_\l \right\} \label{pEffective} \\
 \theta_\l &\quad& \text{such that} \quad \la n \ra_{P_\text{Eff}(\,n\,|\, \l,\g,\d\,)} = \lambda~.
\end{eqnarray}
where as before $\theta_\l$ is not a free parameter, but a uniquely defined function of $\l$ with parameters $\g$ and $\d$.
The probability distribution (\ref{pEffective}) belongs to the class of weighted Poisson distribution and its mathematical properties have been already studied elsewhere \cite{Castillo05}.

We infer $\g$ and $\d$ through a log-likelihood maximization on the OFF and ON cell separately.
For the OFF population we obtain similar values than the Second-Order model (fig.~\ref{fig_3}A).
By contrast, for the ON population the Effective model takes advantage of the additional free parameter and uses it to improve its performance.
We obtained the values 
$\g^*_\text{ON} = -0.52$ and $\d^*_\text{ON} = 0.15$, while the equivalent parameter values in the Second-Order model are $f_\text{ON}^*-f_\text{ON}^{*2}= 0.15$ and $f_\text{ON}^{*2}/2 = 0.02$.

This Effective model is therefore a simple model able to describe accurately sub-Poisson neural variability in several cases, with only two parameters.


\section{Benchmark of proposed models}
\label{comparison}

The Effective model outperforms a Poisson model at predicting the empirical spike count variance.
In this section we compare its performance with two other spike count models proposed in the literature.
The Generalize Count (Gen.Count) model \cite{Gao15} can be seen as a generalization of our Effective model and thus offers a larger flexibility to model the spike count statistics.
This however comes at the price of introducing more parameters to fit, and could potentially lead to overfitting. 
The Conwey-Maxwell-Poisson (COMP) model \cite{Sellers12} has been proposed to account for both under- and over-dispersed mean-variance relation \citep{Stevenson16}.
COMP is a one-parameter extension of Poisson, which differs from our Effective model, but it is still a particular case of Gen.Count (see Methods for details).
For the sake of completeness we also compare the Refractory model, $P_\Th$ of Eq. (\ref{fullEquilRefr}) and its second order expansion, the Second-Order model, $P^\text{2nd}_\Th$ of Eq.(\ref{pSecond}).

\begin{figure}[ht]
\begin{center}
\includegraphics[clip=true,keepaspectratio,angle=-0,width=1.0\columnwidth]{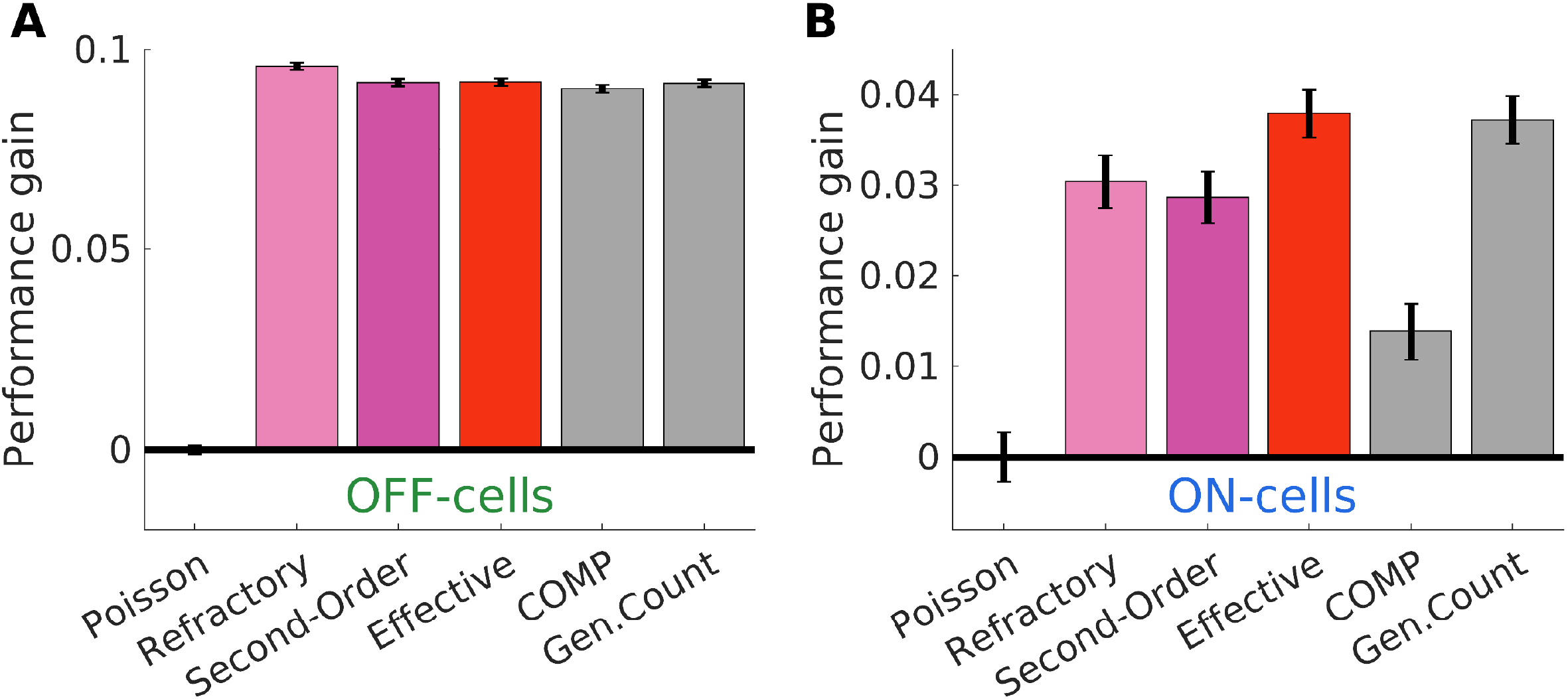}
\end{center}
\caption{\textbf{Second-Order (eq. \ref{pSecond}) and Effective (eq. \ref{pEffective}) model performance compared with known spike count models: COMP (eq. \ref{pCOMP}) and Generalized Count (eq. \ref{pGC})}
Log-likelihood improvement over the Poisson for several models, all learned on a separate training set, for OFF cells ({\bf A}) and ON cells ({\bf B}). 
To estimate log-likelihoods only time-bins with $\l_i(t)>0.3$ are considered (see text).
}
\label{fig_4}
\end{figure}
We compare the performance of different models as improvement over the Poisson log-likelihood (fig.~\ref{fig_4}).
In order to focus on transient with high firing rates, Log-likelihood are estimated on the time-bins of the testing set with $\l_i(t)>0.3$. 
For the OFF population (fig.~\ref{fig_4}A), all models outperform Poisson and have similar performance. 
This is probably because OFF cells show rather small firing rate and emit rarely more than two spikes in the same time-bin.
The addition of one parameter (with respect to Poisson) is thus enough for accurately modeling the spike count statistics.
However, for the ON population (fig.~\ref{fig_4}B), while all the considered models outperform Poisson, Effective and Gen.Count show the largest improvement.
Remarkably, the one-parameter Refractory and Second-Order models show very high performance as well, despite the first being learned by fitting the mean-variance relation of the spike count rather than by maximizing the likelihood.
Also in this case, the larger flexibility of the Gen.Count model does not bring an improvement.

\begin{figure}[ht]
\begin{center}
\includegraphics[clip=true,keepaspectratio,angle=-0,width=1.0\columnwidth]{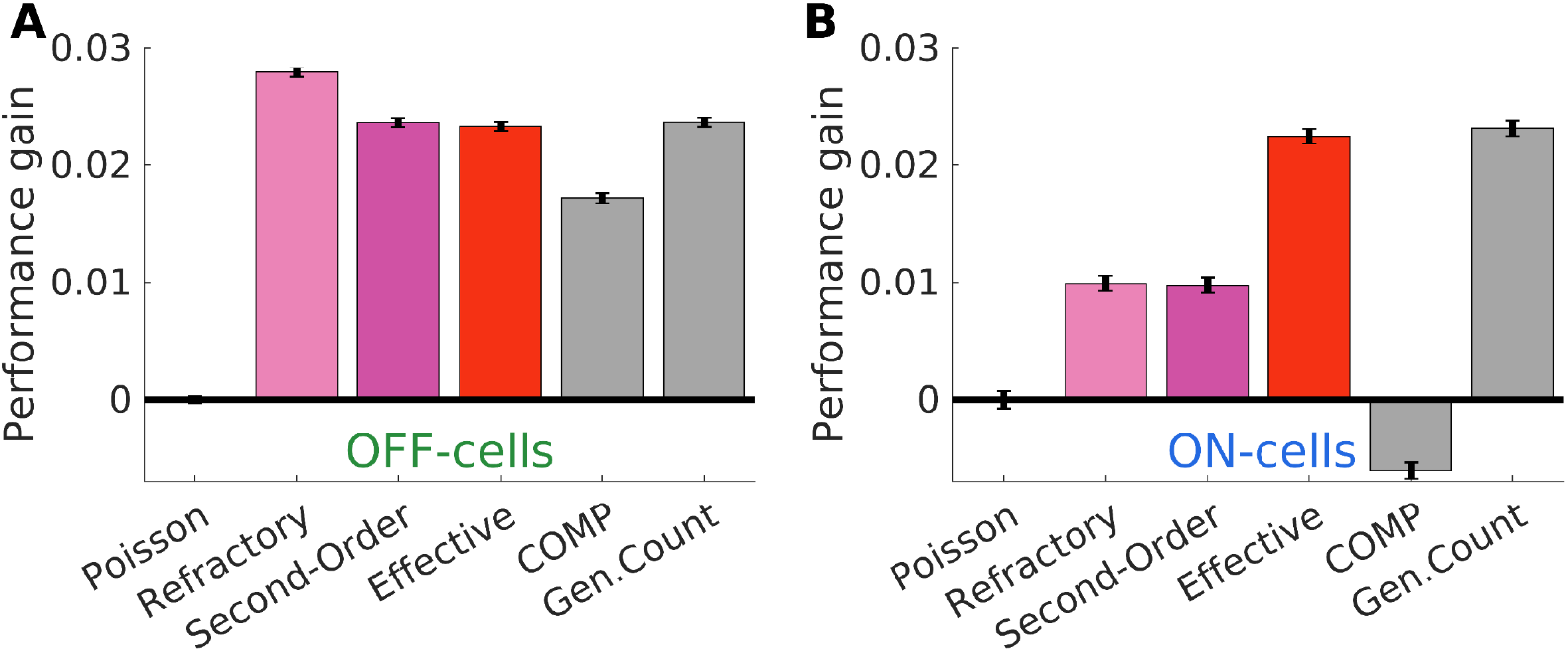}
\end{center}
\caption{\textbf{Performance of different models on response to checkerboard stimulation}
Same as fig.\ref{fig_4} but the testing set was the response to a repeated checkerboard stimulation}
\label{fig_5}
\end{figure}
To test if these models can generalize to other stimulus statistics, after learning the parameters of the model on the responses to the moving bars stimuli, we test them on the responses of the same cells to a repeated sequence of checkerboard stimulus. We find that neurons had also sub-Poisson behavior in response to this stimulus, with spike count variances smaller than the corresponding means, very similar to fig.~\ref{fig_1}C\&D.
Fig.~\ref{fig_5} shows the log-likelihood improvement over the Poisson model.
All models performed better than Poisson, except for the COMP model that for the ON cells does not seem to generalize to other stimuli.
In particular, for OFF cell we found that the Refractory model has the best performance, just a bit larger than the Second-Order and the Effective model (both have very similar performances, see previous section).
For the ON cells Effective and Gen.Count models show the best performance.
Also in this case, the larger flexibility of Gen.Count does not bring a performance improvement. 
The Effective model is thus a simple model to describe deviation from Poisson statistics with only two parameters. 
It works as well as the most general model, Gen.Count, to describe the sub-Poisson variability, but with less parameters. 
It also performs better than other models with a low number of parameters.


\section{Impact of noise distribution on information transfer}
\label{information}

The ability of neurons to transmit information is limited by their variability \cite{Movshon00}.
If neurons are less reliable, or equivalently have larger variance, their capacity to transmit information should be significantly decreased.
To properly estimate the amount of information transmitted by a neuron, a model should reproduce  such variability. 
To test for this, we quantify the amount of stimulus information encoded by emitted spikes as the mutual information $MI$ between the spike count $n$ and its mean $\l$:
\begin{equation}
MI(n,\l) = H[~ P(n) ~] - \big\la ~H[~ P(n|\l)~ ] ~\big\ra_\l \label{eqMI}
\end{equation}
where $H$ is the entropy function and $P(n)$ is the distribution of the spike count without conditioning on the mean over repetitions.
$\la \dots \ra_\l$ is the average over the observed mean spike count and $P(n|\l)$ is the distribution of $n$ at fixed $\l$, either empirically estimated from the repeated data, or estimated with the model.
To avoid under-sampling issues typical of time bin with low activity and due to the finite number of stimulus repetitions, we restrict the averages to all cells and time-bins with $\l_i(t) > 0.1$.

\begin{figure}[ht]
\begin{center}
\includegraphics[clip=true,keepaspectratio,angle=-0,width=1.0\columnwidth]{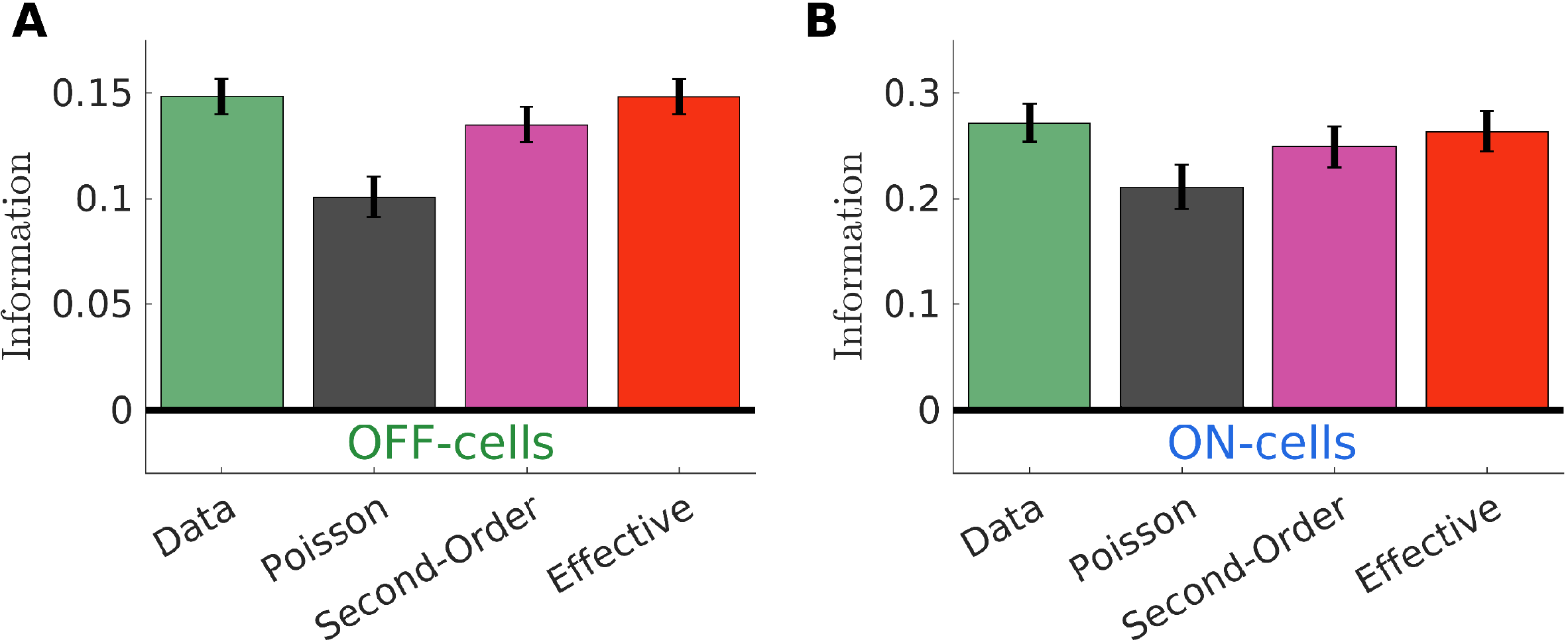}
\end{center}
\caption{\textbf{Second-Order and Effective model predicts empirical information transmission}
Empirical estimation of the mutual information is compared with prediction for Poisson (black), Second-Order (pink) and Effective (red) models.
using Poisson as noise distribution for the spike count leads to a strong under-estimation, biasing the prediction of information transfer of stimulus processing models.
Error-bars are standard deviation of the mean over $\l$, see Eq.~(\ref{eqMI} ).
}
\label{fig_MI}
\end{figure}

In Fig.~\ref{fig_MI}, for both OFF and ON population, Poisson largely under-estimates mutual information, whereas the Effective model predicts well the value of mutual information. 
Our model can thus be used to correctly estimate the mutual information thanks to its accurate prediction of the mean-variance relation.


\section{Improving stimulus processing models}
\label{NLmodel}

The Effective model, Eq.~(\ref{pEffective}) describes efficiently the  relation between mean and variance and predicts well the amount of information transmitted by a neuron. 
Here we show how it can easily be plugged to any stimulus processing model like the Linear-Nonlinear (LN) model, instead of a Poisson process. 
To estimate the parameters of LN models, a classical approach is to assume a Poisson process for spike generation, and then to maximize the likelihood of the spiking data with respect to the parameters of the model. 
The major advantage of this method is that the gradient of the likelihood has a very simple form that allows for iterative log-likelihood maximization. 
Here we show that using the Effective model also leads to a tractable form of the log-likelihood gradient, which can similarly be used for iterative optimization, but with the added advantage that the mean-variance is accurately reproduced.

In general, a stimulus encoding model is defined by a series of computations  - parametrized by parameters $\psi$ - that takes the past stimulus $S_t$ as input and provides a prediction $\hat \lambda_\psi(S_t)$ for the spike count mean as a function of time $t$.
Only at the last stage a stochastic spike counter $P$ is introduced to predict the number of spikes $n(t)$ emitted in the time-bin $t$:
\begin{equation}
{  S_t}~ \rightarrow ~ \hat \lambda_\psi(S_t)~ \rightarrow ~ n(t) \sim P \big(\,n(t) \, \big| \, \hat \lambda_\psi(S_t) \,\big) \label{stimProsModel}
\end{equation}
where for example, $P$ is a Poisson distribution.
The classical example for this is the Linear-Nonlinear-Poisson (LNP) model \cite{Chichilnisky01}, but many generalizations have been proposed, especially for the retina \cite{Mcfarland13,Mcintosh16,Deny17}.

One of the major advantage of using a Poisson spike counter $P_\text{Pois}$ is that it allows for a straightforward optimization of the model parameters ${\psi}$ \cite{Chichilnisky01,Mcfarland13}.
Thanks to the explicit expression of the log-likelihood $\ell({  \psi})$ (see Methods), the log-likelihood gradient ${  \nabla_\psi} \ell({  \psi}) \equiv d \ell({  \psi})  /  d {  \psi}$ takes a very simple form:
\begin{equation}
\!\!\!\!\!\!{  \nabla_\psi} \ell({  \psi}) = \frac{1}{T}  \sum_t \left.  \frac{d\, \log P_\text{Pois} \big(\,n(t) \, \big|\,  \lambda \,\big) }{d  \l}\right|_{  \lambda =  \hat \l_{ \psi}( S_t )} \!\!\!\!\!\!\!\!\!\!\!\!\!\!\! {  \nabla_\psi} \hat \l_{  \psi} ( {  S_t} ) =  \frac{1}{T}  \sum_t  \left( \frac{n(t)  - \hat \l_\psi (S_t)}{\hat \l_\psi (S_t) }  \right) {  \nabla_\psi} \hat \l_{  \psi} ( {  S_t} )~,\label{DlogL_Pois}
\end{equation}
Once ${  \nabla_\psi} \hat \l_{  \psi} ( {  S_t} )$ is evaluated, (\ref{DlogL_Pois}) allows for iterative log-likelihood maximization.

In the previous section we showed that Effective model outperform Poisson as spike counter distribution.
Its particular structure, Eq.~(\ref{pEffective}), allows an easy estimation of the log-likelihood gradient (see Methods for the calculation details):
\begin{equation}
{  \nabla_\psi} \ell({  \psi}) =  \frac{1}{T}  \sum_t  \left(  \frac{ n(t) - \hat \l_\psi(S_t)  }{ V^\text{Eff}_{ \hat \l_\psi(S_t)} }   \right) {  \nabla_\psi} \hat \l_{  \psi} ( {  S_t} ) \label{DlogL_2nd}
\end{equation}
With respect to the Poisson case, the inference of the parameters $\psi$ requires only the variance $V^\text{Eff}_{ \l}$ as a function of $ \l$, which depends only on the noise model and can be easily estimated from it before running the inference (see Methods for more details).

\begin{figure}[ht]
\begin{center}
\includegraphics[clip=true,keepaspectratio,angle=-0,width=1.0\columnwidth]{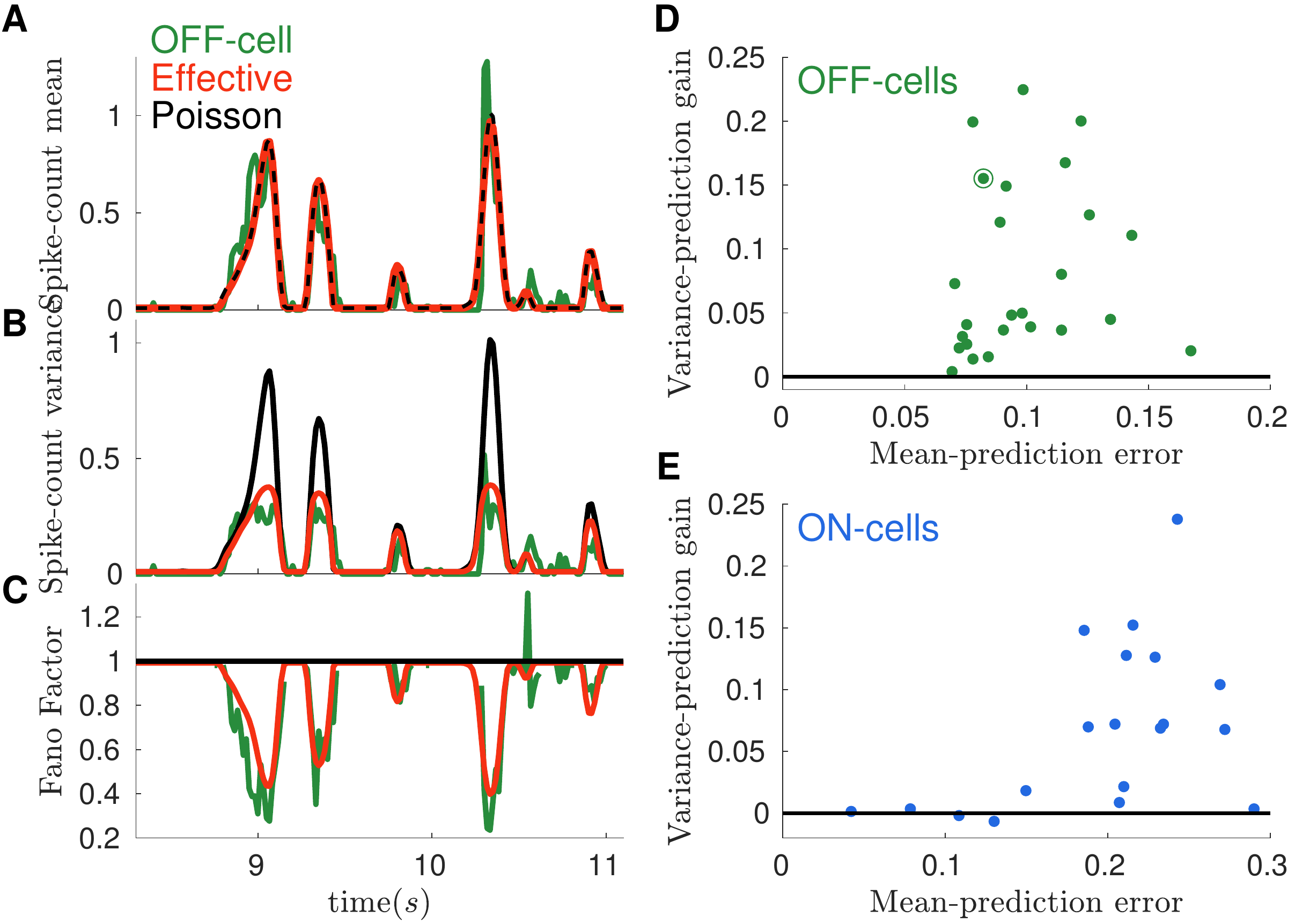}
\end{center}
\caption{\textbf{Effective models improve performance of stimulus processing models}
{\bf A}) PSTH for an example OFF cell (Green) is compared with predictions by the LN$^{2}$ models equipped with Poisson (black) or Effective (red) spike counter.
{\bf B}) As in panel {\bf A}, but for the  spike count variance.  Poisson model largely overestimates the empirical variance during transients of high activity.
{\bf C}) Fano Factor  as function of time: Effective  model (red) accounts for the empirical behavior (green). Poisson prediction (black constant line) does not. 
The empirical trace is not defined when the mean spike count is zero.
{\bf D}) Improvement with respect to Poisson in the mean square error (MSE) between empirical and model variance,  plotted as a function of the  MSE between empirical and model mean spike count. Each point represents a cell of the OFF population.
Circled point refers to the example cell in panels {\bf A}-{\bf C}.
{\bf E}) Same as {\bf D}, but for the Effective model applied on the ON population.
}
\label{fig_6}
\end{figure}

As in \cite{Deny17} and for both OFF and ON populations, we infer a Linear-Nonlinear-Linear-Nonlinear-Poisson (LN$^2$-Pois) model to predict the average cell response to the two bar stimulation.
This is a stimulus processing model composed by a cascade of two layers of linear-filtering\&Nonlinear-transformation followed by a Poisson spike counter.
Similarly, we infer a two-layer cascade Effective model (LN$^2$Eff)  to predict the average cell response of OFF and ON cells respectively. 
LN$^2$-Eff differs from LN$^2$-Pois in the noise generator, either Effective, see Eq.~(\ref{pEffective}), either Poisson.
The two models show very similar prediction for the mean spike activity (fig.~\ref{fig_6}) while LN$^2$-Pois largely over-estimate the spike count variance (fig.~\ref{fig_6}B).
LN$^2$-Eff predicts also well the Fano Factor estimated over time
(fig.~\ref{fig_6}C).
For most of the OFF cells, replacing Poisson by the Effective model leads to a significant performance improvement when trying to predict the variance (fig.~\ref{fig_6}D). 
For OFF cells, the Second-Order model, Eq.~\ref{pSecond}, performs as well as the Effective model.
The Effective model also predicts well variance for ON cells (fig.~\ref{fig_6}E). 
Our Effective model can therefore be plugged in encoding models to describe accurately the variance of the spike count over time.


\section*{Discussion}

We have shown that a simple model taking into account refractoriness in the spiking process explains most of the deviation from Poisson that we observe in the spike count statistics. The model has only two parameters, can easily be plugged into any encoding model to be fitted to sensory recordings.
It allows for an accurate estimation of the relation between the mean spike count and its variance, but also of the amount of information carried by individual neurons. 
The form of this model is inspired by the regularity imposed on spike trains by the refractory period. However, it can potentially work for data with other sources of regularity.
In the retina, this model works for two different types of cells.
Thanks to its simplicity and generality, this model could potentially be used to account for mean-variance relation in neurons recorded in other sensory areas \cite{Gur97,Kara00,Barberini01,Deweese03,Maimon09}, even if their neural variability is not solely determined by the refractory period in the spike generation of the neuron. Previous works \citep{Movshon00,Kara00} suggested that the refractory period present in the phenomenological model may reflect refractoriness at any stage in the circuit and may not directly correspond to the refractoriness of the recorded cell.  


We have compared our model with others already proposed in the literature: the Conwey-Maxwell-Poisson (COMP) \citep{Stevenson16} and the Generalized Count (Gen.Count) \cite{Gao15}.
We found the COMP model was quite inefficient at fitting our data. 
The Gen.Count model can be considered as encompassing a much larger class of possible models, and includes ours as a special case.
However, this comes at the cost of having many more parameters to fit, which can lead to overfitting in some cases, as we have shown in the results.
Moreover, in our data, these additional parameters did not allow improving the performance of the model.
The model we propose has only two parameters, which makes it easy to fit and usable in many cases, even when the amount of data is limited.
It will be interesting to compare our effective models to the Gen.Count model in other sensory structures. 

The relation between mean and variance of the spike count strongly depends on the bin size chosen to bin the cell response.
For very long windows, we cannot assume constant firing rate, and we observe a superimposition of many different rates. In this case, and even for the retina, cells show much larger Fano Factor \cite{Kara00}, that can also exceed unity.
In other systems where some of the sources of variability are not controlled, a similar heterogeneity of firing rates can be observed also because the uncontrolled source will impose a different firing rate at each trial.  
Several solutions have been proposed to model this over dispersion \cite{Scott12,Goris14,Charles17}. 
For very small time bins the number of emitted spikes rarely exceeds unity, so that the spike count becomes a binary variable with fixed mean-variance relation $V = \l(1-\l)$.
However, neuron refractoriness plays an important role and the activity in consecutive small time-bins are strongly correlated, for example cells never spike in consecutive time-bins: $n_i(t)=0$ if $n_i(t-1) =1$.
In this case a common solution is to introduce a spike history filter \cite{Pillow08} that models the spike probability with a dependence on the past activity. 
We have chosen a bin size between these two extremes, such that the spike count is not a binary variable, but the firing rate stays roughly constant within a single time bin and the dependence between consecutive time bins is relatively weak. The advantage of this choice of bin size is that the number of parameters needed to describe the spike regularity is small: only two parameters.
The bin size chosen also corresponds to the timescale of 
the retinal code \cite{Berry98b}.


\section*{Methods}

\textbf{Equilibrium Poisson process with absolute refractory period}
A Poisson process with firing rate $\Theta(u-\t) r$, where $\Theta$ is the Heaviside step, function of the time from previous spike,  has inter-spike interval distribution given by Eq. (\ref{ISI_refr}).
For such process it is possible to compute the probability distribution $P_\Th\big( n \big| r ,\t,\Dt \big)=P_\Th\big( n \big| \nu ,f \big) $, where $f = \t/\Delta t$ and  $\nu = r \Delta t$,  of the number a spikes emitted in a time window $\Delta t$, whose starting point is chosen as random \cite{Muller73,Muller74} (see SI for a derivation):
\begin{eqnarray}
P_\Th\big( n \big| \nu ,f \big) &=& \frac{1}{1+\nu f}\left[\Phi(n) +  \Theta(n^\text{Max}-2-n)   \sum_{j=0}^n (n+1-j) \frac{\nu^j (1-(n+1)f)^j e^{-\nu (1-(n+1)f) }}{j!}  \right. \nonumber \\
&& -2  \Theta(n^\text{Max}-n-1)  \sum_{j=0}^{n-1} (n-j)  \frac{\nu^j (1-n f)^j e^{-\nu (1-n f) }}{j!}  \nonumber \\
&& + \left. \Theta(n^\text{Max}-n)   \sum_{j=0}^{n-2} (n-1-j)  \frac{\nu^j (1-(n-1)f)^j e^{-\nu (1-(n-1)f) }}{j!} \right] \label{fullEquilRefr}\\
 \Phi(n) &\equiv&  \left\{   
\begin{array}{lcc}
0 &~,~~& n \leq n^\text{Max}-2 \\
n^\text{Max} (1+\nu f) -\nu &~,~~& n = n^\text{Max}-1 \\
\nu - (n^\text{Max}-1) (1+\nu f) &~,~~& n = n^\text{Max}
\end{array}
\right.
\end{eqnarray}
where $n^\text{Max}$ is the smallest integer larger than $\Delta t / \t$ and we used the convention $\Theta(0)=1$.
The distribution (\ref{fullEquilRefr}) has expected value given by:
\begin{equation}
\text{E}_\Th(n|\nu,f) = \frac{\nu}{1+\nu f} \label{meanRefr}
\end{equation}
 and exact variance \cite{Muller74}:
\begin{equation}
\text{V}_\Th(n|\nu,f) = \frac{2 \sum_{n=0}^{n^\text{Max}-1} \left[ \nu(1 - nf) - n + \sum_{j=0}^{n-1}  (n-j)  \frac{\nu^j (1-n f)^j e^{-\nu (1-n f) }}{j!} \right] -\nu - \frac{\nu^2}{1+\nu f}  }{1+\nu f}  \label{varRefr}
\end{equation}

\textbf{Inference of the absolute refractory period model}
To infer the value of $f$ for the absolute refractory model, see Eq.~(\ref{fullEquilRefr}) we perform a mean square error (MSE) minimization of the mean-variance relation:
\begin{eqnarray}
MSE &\equiv& \sum_{i,t} \big[~ V_i(t)  - \mathrm{Var}_\Th(n|\nu_i(t),f)~\big]^2 \\
\nu_i(t) &=& \frac{\l_i(t)}{1- \l_i(t) f}
\end{eqnarray} 
where $\l_i(t)$ and $V_i(t)$ are the empirical mean and variance of $n$ for the cell $i$ in the time-bin $t$ of the training set.
We used the exact expression (\ref{varRefr}) for $\mathrm{Var}_\Th(n|\nu_i(t),f)$ although an approximated results can be obtained by the simpler asymptotic expression \cite{Muller74}:
\begin{equation}
\text{V}_\Th(n|\nu,f) \approx \frac{\nu}{(1+\nu f)^3}
\end{equation}

\textbf{Inference of the Second-Order refractory model}
To infer the value of $f$ for the Second-Order refractory model, see Eq.~(\ref{pSecond}) we perform log-likelihood maximization with a steepest descent algorithm.
Thanks to exponential form of  $P_\Th^\text{2nd}(n|\l,f)$, the derivative of the  log-likelihood ($\ell$) with respect to $f$ takes a simple form:
\begin{eqnarray}
\ell &\equiv& \sum_{i,t}  \log P_\Th^\text{2nd}(n|\l_i(t),f) \\
\frac{d\, \ell}{d\,f}&=& \sum_{i,t}  (2f-1) \left(\la n_i(t)^2 \ra_\text{data} - \la n_i(t)^2 \ra_{\text{model}_i(t)}  \right) - f \left(\la n_i(t)^3 \ra_\text{data} - \la n_i(t)^3 \ra_{\text{model}_i(t)}  \right)~,
\end{eqnarray}
where $\la \cdot \ra_{\text{model}_i(t)}$ means average with the model distribution $P_\Th^\text{2nd}(n|\l=\l_i(t),f)$.
At each iteration, we update the value of the parameter using the log-likelihood gradient and we adjust the function $\th_\l$ accordingly.

\textbf{Inference of the Effective model}
To infer the value of $\g$ and $\d$ for the Effective model, see Eq.~(\ref{pEffective}) we perform log-likelihood maximization with Newton method, where at each iteration we update the values of the parameters using the log-likelihood gradient:
\begin{eqnarray}
 \sum_{i,t} \frac{d}{d~\g} \log P_\text{Eff}(n|\l_i(t),f) &=& \la n_i(t)^2 \ra_\text{data} - \la n_i(t)^2 \ra_{\text{model}_i(t)} \\
 \sum_{i,t} \frac{d}{d~\d} \log P_\text{Eff}(n|\l_i(t),f) &=& \la n_i(t)^3 \ra_\text{data} - \la n_i(t)^3 \ra_{\text{model}_i(t)}
\end{eqnarray}
and the Hessian:
\begin{equation}
\mathcal{H} = -  \left(  
\begin{array}{c c}
 \la n_i(t)^2 \ra_{\text{model}_i(t)} - \l_i(t)^2 &  \la n_i(t)^3 \ra_{\text{model}_i(t)} - \l_i(t) \la n_i(t)^2 \ra_{\text{model}_i(t)} \\
 \la n_i(t)^3 \ra_{\text{model}_i(t)} - \l_i(t) \la n_i(t)^2 \ra_{\text{model}_i(t)} &  \la n_i(t)^4 \ra_{\text{model}_i(t)} -  \la n_i(t)^2 \ra_{\text{model}_i(t)}^2
\end{array}
\right)
\end{equation}
where now $\la \cdot \ra_{\text{model}_i(t)}$ means average with the model distribution $P_\text{Eff}(n|\l=\l_i(t),\g,\d)$.
After updating the parameter, we adjust the function $\th_\l$ and iterate until convergence.

\textbf{Generalized Counter model}
In Ref \citep{Gao15}, the authors define a Generalized Counter (Gen.Count) distribution, that in our notation and framework reads:
 \begin{equation}
P_\text{Gen.Count}(~n~|~ \l ~) = \exp\left\{  \theta_\l[G] ~n + G[n] - \log n! -\log Z_{\l}[G] \right\}\label{pGC}
\end{equation}
where $G[n]$ is a generic real function defined on the non-negative integers $n \in [0,\infty]$.
To better characterize its $\l$ dependence, we have rewritten Gen.Count introducing the proper $\th_\l[G]$ function.

For $G[n] = -\g n^2-\d n^3$  the distribution (\ref{pGC}) reduces to Effective model, see Eq. (\ref{pEffective}).
The Gen.Count model is thus a generalization of our model, and as such is potentially more flexible in modeling the spike count statistics. 
This however comes at the price of introducing more parameters to fit. 
Specifically, in practical application one need to define $G[n]$ up to an $n^\text{Max}$, that is $n^\text{Max}+1$ free parameters \cite{Gao15}, but the model is invariant for $G[n] \to G[n]+const.$ and $G[n] \to G[n] +c n$.
Gen.Count has thus $n^\text{Max}-1$ free parameters: if $n^\text{Max}=1,2,3$ then Gen.Count is equivalent to, respectively Poisson, Effective with $\d=0$ and Effective models. 
Otherwise is offers a potentially interesting way to generalize our cubic model at the price of inferring the parameters $G[n]$ for large values of $n$
Thanks to its exponential form, we easily infer the Gen.Count parameters with log-likelihood maximization and we have set $n^\text{Max} = 4$ and $5$ for, respectively $OFF$ and $ON$ populations.

\textbf{Conwey-Maxwell-Poisson model}
In Ref. \citep{Stevenson16} the Conwey-Maxwell-Poisson (COMP) model \cite{Sellers12} has been proposed to account for both under- and over-dispersed mean-variance relation.
In our notation, the COMP model reads:
 \begin{equation}
P_\text{COMP}(~n~|~ \l ~) = \exp\left\{  \theta_\l[\eta]~  n  - \eta \log n! -\log Z_\l[\eta] \right\} \label{pCOMP}
\end{equation}
For $\eta=1$ the COMP reduces to the Poisson model, whereas for $G[n] = - (\eta-1) \log n!$ the Generalized Count reproduces the COMP.
Also for the COMP, the exponential form allows for log-likelihood maximization.

\textbf{Equipping stimulus processing model with Second-Order or Effective noise term}
In this appendix we detail the calculation for computing the log-likelihood gradient of Eq.~(\ref{DlogL_2nd}).
For a general stimulus processing model, see Eq.~(\ref{stimProsModel}), equipped with noise term $P$ the log-likelihood reads:
\begin{equation}
\ell( \psi) = \frac{1}{T} \sum_t \log P \big(\, n(t) \, \big| \, \hat \lambda_\psi(S_t) \,\big) 
\end{equation}
where the summation runs over the duration of the training set.
Consequently the log-likelihood gradient $\nabla_\psi \ell(\psi)$:
\begin{equation}
\nabla_\psi \ell(\psi) = \frac{1}{T}  \sum_t \left.  \frac{d\,  \log P \big(\,n(t) \, \big|\, \lambda \,\big)}{d \l} \,\right|_{\l = \hat \lambda_\psi(S_t)}  \nabla_\psi \hat \lambda_\psi(S_t) ~.\label{app:logLgrad}
\end{equation}
To estimate $\nabla_\psi \ell( \psi) $ we thus need to compute the derivative of the log-probability with respect to the mean spike count.
If $P$ is the Poisson distribution $P_\text{Pois}$, then:
\begin{equation}
 \frac{d}{d  \l}  \log P_\text{Pois} \big(\,n(t) \, \big|\, \lambda \,\big)  =  \frac{d}{d  \l} \Big( n(t) \log  \lambda - \lambda - \log n(t)!   \Big) = \frac{n(t)  -  \l }{ \l }
~. \label{app:DlogL_Pois}
\end{equation}
If $P$ is instead the Effective model:
\begin{eqnarray}
\frac{d}{d  \l}  \log P_\text{Eff} \big(\,n(t) \, \big|\, \lambda \,\big) &=&\frac{d}{d \th_{\l}}  \Big(   \log P_\text{Eff} \big( \,n(t) \, \big| \, \lambda \,\big) \, \Big) ~  \frac{d}{d  \l}   \th_{ \l} \\
 &=& \left(  n(t) -  \l   \right)  \left(\frac{d}{d \th_{ \l} }  \l \right)^{-1} =  \frac{ n(t) - \l  }{ V^\text{Eff}_{ \l} } ~.\label{app:DlogL_Eff}
\end{eqnarray}
The very same expression, with  replaced $V_{ \l}^\text{Eff}$ by $V^\text{2nd}_{\l}$, holds for the Second-Order model.
Note that (\ref{app:DlogL_Pois}) has the same form of (\ref{app:DlogL_Eff}) because for Poisson $V^\text{Pois}_{\l}(t) =  \l$.
Eq. (\ref{DlogL_2nd}) is in fact a general result for spike counter $P$ belonging to the exponential family.
Note that  $V_{\l}^\text{Eff}$ and $V^\text{2nd}_{ \l}$ are properties of the noise distribution (respectively $P_\text{Eff}$ and $P^\text{2nd}_\Th$).
These functions can be numerically estimated before running the inference of the stimulus processing model, for example, by computing their values for several values of $\l$ and then interpolating with a cubic spline.


\subsection*{Acknowledgments}
We like to thank R. Brette, M. Chalk, C. Gardella, B. Telenczuk and M. di Volo for useful discussion.
This work was supported by ANR TRAJECTORY, the French State program Investissements d'Avenir managed by the Agence Nationale de la Recherche [LIFESENSES: ANR-10-LABX-65], a EC grant from the Human Brain Project (FP7-604102)), NIH grant U01NS090501 and AVIESAN-UNADEV grant to OM

\appendix
\section{Equilibrium Poisson process with absolute refractory period}
\label{app:refractoryModel}

Here we provide a sketch of the derivation to obtain the complete expression (\ref{fullEquilRefr}).
We are interested in computing the probability distribution $P_\Th\big( n \big|  r ,\t,\Delta t \big)$ of the number a spikes emitted in a time window $\Delta t$, whose starting point is chosen as random, when the inter-spike interval distribution is:
\begin{equation}
\rho_\Th(t) = \Theta(t-\tau) r e^{- r ~ (t - \tau)} ~.
\end{equation}
$P_\Th$ can be expressed as difference between its cumulative distribution:
\begin{eqnarray}
C(n|\Dt) &\equiv& \sum_{k=n}^{\infty} P_\Th\big( k \big| \Delta t, r ,\t \big) \\
P_\Th\big( n \big| \Delta t, r ,\t \big) &=& C(n|\Dt) - C(n+1|\Dt)~.
\end{eqnarray}
Because $C(n|\Dt)$ is the probability of having at least $n$ spikes in the time-bin $\Dt$ it can be computed as:
\begin{equation}
C(n|\Dt) = \int_0^{\D t} dt~ [\rho^E_\Th \star \rho_\Th^{\star (n-1)}](t) \label{cumulRefr}
\end{equation}
where $\star$ is the convolution symbol and $(\cdot)^{\star n}$ means $n$-times self convolution. $\rho^E_\Th$ is the distribution of the first spike when the beginning of the time-bin is chosen at random (equilibrium process) and its distribution can be computed as \cite{Muller73}
\begin{eqnarray}
\rho_\Th^E(t) &=& \frac{ \int_t^\infty  dt'~ \rho_\Th(t') }{ \int_0^\infty  dt'~ t'~\rho_\Th(t') } = r ~\frac{e^{- r \max(t- \t,0)}}{1 + r \tau} \nonumber \\
&=& r ~\frac{\Theta(t)\Theta(\t-t) +  \Theta(t-\t)e^{- r (t- \t)}}{1 + r \tau} = \frac{1}{1 + r \tau}\left( r \mathcal{I}_{[0,\t]}(t) + \rho_\Th(t) \right) ~,
\end{eqnarray}
where $\mathcal{I}_{[a,b]}(t) = 1$ if $t \in [a,b]$ and zero otherwise.
Thanks to the explicit decomposition of $\rho_\Th^E$ we have:
\begin{equation}
\rho^E_\Th \star \rho_\Th^{\star (n-1)} =  \frac{1}{1 + r \tau}\left( r \mathcal{I}_{[0,\t]} \star \rho_\Th^{\star (n-1)} + \rho_\Th^{\star n} \right)~.
\end{equation}
In order to estimate the $\rho_\Th^{\star n}$ we introduce the Laplace transform, which for a generic function $h(t)$ reads:
\begin{equation}
\mathcal{L}[h(t)](s) = \int_0^\infty dt~ e^{- s t} h(t)~,
\end{equation}
and use it to get read of the multiple convolutions:
\begin{eqnarray}
\rho_\Th^{\star n} &=&  \mathcal{L}^{-1}\left[~\mathcal{L}[~ \rho_\Th^{\star n}~]~   \right] = \mathcal{L}^{-1}\left[~\mathcal{L}[~ \rho_\Th~]^n  ~ \right] =\mathcal{L}^{-1}\left[ ~\left(\frac{r}{r + s}e^{-s \t} \right)^{n} ~ \right] \nonumber \\
&=&  \frac{r^n (t-n \t)^{(n-1)} e^{-r (t-n\t) }}{(n-1)!} \Th(t-n\t)  \equiv   \g[n,r](t-n\t) \label{eq:gamma}
\end{eqnarray}
where we have introduced  the Gamma distribution $\g[n,r](t)\equiv \Th(t) \,r^n t^{(n-1)} e^{-r}/(n-1)!$.
From (\ref{cumulRefr})  it follows:
\begin{eqnarray}
C(n,\Dt) &=& \frac{1}{1 + r \tau}  \left( ~\int_0^{\Dt} dt~ \int dt'~  \mathcal{I}_{[0,\t]}(t-t') \g[n-1,r](t'-(n-1)\t) \right. \nonumber \\
&&\qquad \qquad +  \left.  ~\int_0^{\Dt} dt~ \g[n,r](t- n\t)  ~ \right)~.
\end{eqnarray}
$C(n|\Dt)$ can be computed by integrating  several times the gamma distribution using the following relation:
\begin{equation}
 \frac{1}{(n-1)!}\int_0^a dt~ t^{(n-1)}e^{-t} \Th(t) = \left( 1 - e^{-a} \sum_{k=0}^{n-1} \frac{a^k}{k!} \right) \Th(a) ~.
\end{equation}
After some algebra \cite{Muller73, Muller74} this calculation provides Eq.~(\ref{fullEquilRefr}).


\section{Small $f$ expansion of the refractory neuron model}
\label{app:expansion}

Here we derive the result (\ref{pOfNalpha}), for which (\ref{pSecond}) is a particular case.
We considered a general model in which the instantaneous spike rate is modulated by a time-dependent factor, $\alpha(u)r$, where $r$ is the spike rate in absence of refractoriness, with $\alpha(u\geq \t)=1$, and $u$ is the time following the previous spike.
In this case the inter-spike interval distribution is:
\begin{equation}
\rho_\a(t) = \nu \a(t) \exp\left\{- \nu \int_0^t \a(t)   \right\}~\Th(t)~,
\end{equation}
and our goal is to expand, for small $\t$, $P_\a(n| \Dt, r,\a)$, the probability of having $n$ spikes within the time-bin $\Dt$.
At first we introduce two  useful quantities:
\begin{eqnarray}
f &\equiv& \frac{1}{\Dt} \int_0^\infty dt~  \big(1-\a(t)\big) \label{eqAt} \\
g &\equiv& \frac{1}{\Dt} \int_0^\infty dt~ t~\big(1- \a(t)\big)  \label{eqBt}
\end{eqnarray}
Such that, if $\a(t) = 1$ for all $t>0$, then $f=g=0$ and we expect to recover the Poisson case, and if $\a(t) = 0$ for all $t<\t$, then $f=\t/\Dt$ and $g=\t^2/\Dt^2/2$ and we expect to recover the absolute refractory case, \text{i.e.} Eq.~(\ref{pSecond}).

Much like the absolute refractory case, we consider the cumulative distribution of $P_\a(n|  r,\a,\Dt,)$:
\begin{equation}
C_\a(n| \Dt) = \int_0^{\Dt} ~ [\rho^E_\a \star \rho_\a^{\star (n-1)}](t) \label{cumulExpans}~,
\end{equation}
where, as before, $\rho^E_\a$ is the distribution of the first spike for an equilibrium process and $\rho_\a^{\star (n-1)}$ is $\rho_\a$ self-convoluted $n-1$ times.
To perform the expansion in small $\t$, we decompose $\rho^E_\a$ and $\rho_\a$ around an exponential distribution:
\begin{eqnarray}
\rho^E_\a(t) &=&   \left(1+\nu^2g \right) \rho(t)  + \d\rho^E_\a(t)  \\
\rho_\a(t) &=& e^{r (1-A)\t} \rho(t) + \d \rho_\a(t)~.
\end{eqnarray}
In the following we will first perform the computation with $\rho(t)$ instead of $\rho^E_\a$, leaving for the end the corrections due to the factor $1+\nu^2 g$ and term $\d\rho^E_\a$.
This is equivalent to consider a \textit{shifted} process \cite{Muller73} where the time-bin start after the end of the last refractory period, instead of the \textit{equilibrium} process we are considering here.
As for $\rho_\Th^{\star n}(t)$, see Eq.~(\ref{eq:gamma}), we can use the Laplace transform to compute $\rho^{\star n}(t) = \g[n,r](t)$.
This allows us to perform the following expansion:
\begin{eqnarray}
[\rho \star \rho_\a^{\star (n-1)}](t) &=& \rho \star \sum_{i=0}^{n-1} {{n-1}\choose{i}} e^{(n-i-1)\nu f} \big[~\rho^{\star (n-1-i)} \star  \d\rho_\a^{\star i}~\big](t) \\
&\approx&  e^{(n-1)\nu f} \g[n,r](t) \nonumber \\ 
&&\qquad + (n-1)e^{(n-2)\nu f} \big[~\g[n-1,r] \star \d\rho_\a~\big](t) \nonumber \\
&&\qquad +{{n-1}\choose{2}} e^{(n-3)\nu f} \big[~\g[n-2,r] \star  \d\rho_\a^{\star 2}~\big](t)~. \label{convolutions}
\end{eqnarray}
To perform the integration of (\ref{convolutions}) we use the following approximation:
\begin{eqnarray}
\int_0^\infty dt~ \d \rho_\a(t)  &\approx& - \nu f -\frac{1}{2} \nu^2f^2+  O(\t^3)  \\
\int_0^\infty dt~ \d \rho_\a^{\star 2}(t) &\approx&   \nu^2 f^2 +  O(\t^3)\\
\int_0^\infty dt~ t~ \d \rho_\a(t) &\approx& -\nu g \Dt +  O(\t^3)~,
 \end{eqnarray}
which after some algebra we obtain for $P^S_\a(n| \nu )$, the distribution for the \textit{shifted} process:
\begin{equation}
P^S_\a(n| \nu ) = \int_0^{\Dt}dt~  \Big[\rho \star \left(  \rho_\a^{\star (n-1)}  - \rho_\a^{\star n}  \right) \big](t)   \propto \exp \left\{ n \log \nu - \log n! + c_1 n +  c_2 n^2 +  c_3 n^3  \right\} \label{pOfNalphaTilde}\\
\end{equation}
with
\begin{eqnarray}
c_1 &=&  (1+ \nu) \left(f -\frac{f^2}{2}\right) - \left( \frac{f^2}{2} -g \right) \\
c_2 &=&  -  f + f^2 - (2+\nu)  \left(\frac{f^2}{2}-g \right) \\
c_3 &=& -f^2+g 
\end{eqnarray}
We need now to account for the full $\rho_\a^{E}(t)$.
Because
\begin{equation}
\int_0^\infty dt~ \d\rho^E_\Th(t)  \approx  - \nu^2 g
\end{equation}
is of the order $\t^2$ we can estimate $P_\a(n| \Dt, r)= P_\a(n| \nu)$ as:
\begin{eqnarray}
P_\a(n| \nu ) &=&  \left(1+\nu^2 g\right) P^S_\a(n|\nu ) + P_\text{Pois}(n-1| \nu) \int_0^\infty dt~ \d\rho^E_0(t) \nonumber\\
&\propto& \exp \left\{ n \log \nu - \log n! + \left(c_1 - \nu g \right) n +  c_2 n^2 +  c_3 n^3  \right\}~.
\end{eqnarray}
which is equivalent to (\ref{pOfNalpha}).


\bibliographystyle{unsrt} 
\bibliography{../../../updatedFileBibliography/Ulisse.bib}

\end{document}